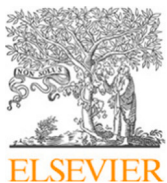
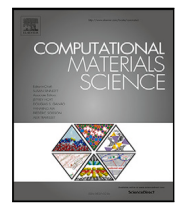

Full length article

# First-principles study of the optical properties of BaMoO$_3$/SrHfO$_3$ hyperbolic metamaterials

Jonathan Gjerde [*], Radi A. Jishi

*Department of Physics and Astronomy, California State University, Los Angeles, CA 90032, USA*

A B S T R A C T

The theoretical applications of hyperbolic metamaterials have generated excitement in multiple fields, particularly in super-resolution optics. In practice, however, the potential of HMMs has been limited by the shortcomings of their constituent parts. Primarily, these are high losses and instabilities in the metal component, which is typically gold or silver, and excessive thickness in both types of layers. In this paper, we use density functional theory to investigate BaMoO$_3$-SrHfO$_3$ metamaterials. We predict that these are hyperbolic throughout the visible range and have the potential to mitigate many of these limitations, including a significant loss reduction. We also examine the accuracy of the effective-medium theory, which is used to predict the bulk optical properties of multilayered HMMs.

## 1. Introduction

Metamaterials are artificial structures that offer the prospect of breaking what have previously been considered the physical limits of optics, such as the diffraction limit. They display unusual optical properties like negative refraction [1]. Among the most promising categories of metamaterials are hyperbolic metamaterials (HMMs). These are materials whose permittivity along the optical axis has opposite sign to that along the in-plane direction [2]. The most important consequence of this is that HMMs support large k-vectors, allowing them to be used in novel devices and applications.

HMMs are anisotropic materials, with symmetry along one spatial direction (taken to be the $z$-direction) different from that along the other two directions. This causes the optical properties of these materials to vary depending on whether light is perpendicular or parallel to the layers, leading to a dielectric tensor with two unique components ($\epsilon_{xx} = \epsilon_{yy} \neq \epsilon_{zz}$). In general, these components will also depend on frequency. The equation relating frequency, dielectric components, and wave vector components is the dispersion relation. A typical uniaxial, anisotropic, nonmagnetic material (magnetic permeability $\mu = 1$) obeys the dispersion relation:

$$\left(\frac{\omega}{c}\right)^2 = \frac{k_x^2 + k_y^2}{\epsilon_{zz}} + \frac{k_z^2}{\epsilon_{xx}} \quad (1)$$

Commonly, both $\epsilon_{xx}$ and $\epsilon_{zz}$ are positive; this results in constant frequency surfaces that are bounded ellipsoids in momentum space. If one, but not both, of $\epsilon_{xx}$ and $\epsilon_{zz}$ is negative, then the dispersion is hyperbolic, and the constant frequency surfaces are no longer bounded (see Fig. 1). If $\epsilon_{zz}$ is positive while $\epsilon_{xx}$ is negative, the surface is a hyperboloid of one sheet, which is called a type 1 HMM. When $\epsilon_{zz}$ is negative while $\epsilon_{xx}$ is positive, the surface is a hyperboloid of two sheets, called a type 2 HMM. In either hyperbolic case, for a given frequency, there is no limit to the magnitude of the wave vector, leading to many of the unique physical properties of HMMs and the potential for new or improved optical devices [3]. Perhaps the most studied of these proposed devices is the hyperlens, which breaks the diffraction limit to focus extremely small details of incoming images [4]. In nanolithography, the hyperlens can also be used for the miniaturization of larger images, also beyond the diffraction limit [5]. Other important applications of HMMs include negative refraction [6], spontaneous emission engineering [7], and thermal emission engineering [8].

There are some naturally occurring materials with hyperbolic dispersion. These include sapphire, bismuth, hexagonal boron nitride, graphite, and some layered transition-metal dichalcogenides [9–12]. Most of these materials are hyperbolic in the infrared frequency range. The operating wavelength of these materials is fixed by their crystal structure and can be only slightly tuned by external factors such as temperature and pressure. Typically, however, HMMs are manufactured by artificially constructing a multilayer or nanowire composite of a metal and a dielectric [13]. In a multilayer material, the metal and dielectric alternate in layers. This insulates the metallic layers from one another, making the material metallic parallel to the layers and insulating perpendicular to them. Since metals have a negative dielectric function for some range of frequencies, $\epsilon_{xx}$ is negative while $\epsilon_{zz}$ is positive, resulting in a type 1 HMM. In the nanowire configuration, metal nanowires are embedded in a dielectric matrix, insulating them from one another and giving a type 2 HMM. The optical properties of these metamaterials can be adjusted by varying the constituent metal and dielectric components, or by varying the volume fraction of the metal or dielectric in the composite system.

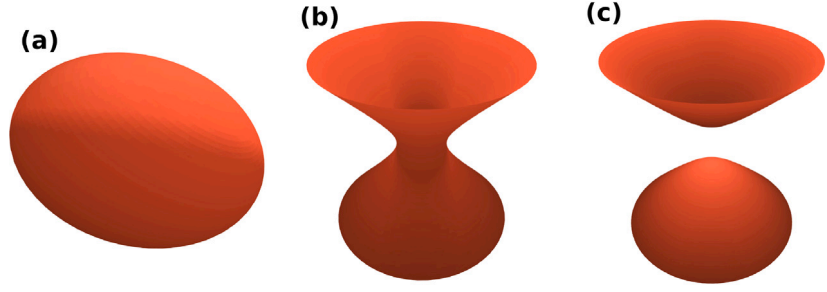

**Fig. 1.** Constant frequency surfaces in momentum space: (a) ellipsoidal ($\epsilon_{xx} > 0$, $\epsilon_{zz} > 0$), (b) type 1 hyperbolic ($\epsilon_{xx} < 0$, $\epsilon_{zz} > 0$), and (c) type 2 hyperbolic ($\epsilon_{xx} > 0$, $\epsilon_{zz} < 0$).

The most widely studied class of HMMs uses a metal oxide (most commonly an aluminum oxide or titanium oxide) as the dielectric and gold or silver as the metal in a multilayer structure [14]. However, the use of gold or silver as the metal component limits these materials' ability to take full advantage of the properties of HMMs, primarily because of large losses [15]. These losses are characterized by the imaginary part of the dielectric function. The use of gold and silver HMMs in devices is also limited by thermal instability, an inability to create extremely thin layers through epitaxial growth, and CMOS incompatibility [16]. All of this highlights the importance of finding alternative materials for HMMs [17]. In this paper, we use density functional theory (DFT) to study the optical properties of a potential new class of multilayer perovskite–perovskite HMMs, and we compare our results with the predictions of the effective-medium theory (EMT). The metal component in the proposed metamaterial is $BaMoO_3$ (BMO), while the dielectric is $SrHfO_3$ (SHO). Both compounds have the ideal cubic perovskite structure, and they have nearly identical lattice parameters.

Perovskite crystals and nanostructures, with chemical formula $ABX_3$, have been widely studied due to their many applications in solar cells, photodetection, and LEDs [18–20]. These applications typically require a very narrow range of band gaps and optical properties. Where perovskites shine is in their ability to adjust these properties to suit the needs of devices by varying the A or B ion, or by partially substituting either of these [21–23]. For example, to increase conductivity, the A site may be partially substituted with lanthanum or praseodymium, while the B site may be substituted with iron, cobalt, or nickel [24]. Altering the conductivity would change the effective mass of the electrons within the material. This, in turn, would affect the intraband contribution to the permittivity, which will be explained in the following section. Other known methods of perovskite band gap tuning, such as the application of pressure [25], may also be able to optimize the optical properties of perovskite-based optoelectronic devices.

Epitaxial growth has been demonstrated for many perovskites. Very thin films of SHO have been successfully grown using pulsed laser deposition [26] as well as molecular beam epitaxy [27]. These methods have not yet been tested for BMO, but the number of perovskites epitaxially grown with these and other methods is constantly increasing [28,29]. It has also been demonstrated that cubic perovskites may be epitaxially grown on top of one another, as long as the lattice mismatch is not too large [30].

This stacked growth and substitution was essential to a previously demonstrated class of perovskite–perovskite HMMs employing $La_xSr_{1-x}TiO_3$ (LSTO) as the metallic component and $SrTiO_3$ (STO) as the dielectric. These HMMs were shown to be hyperbolic in the near- to mid-infrared range and exhibited large Purcell factors [31].

The BMO-SHO metamaterial is likely more thermally stable than gold and silver based HMMs. The melting temperature of SHO is 3200 K [32] and that of BMO is 1791 K [33], somewhat higher than that of gold (1337 K) and silver (1235 K) [34].

Layered metamaterials like those examined in this work are modeled as homogeneous anisotropic materials with an effective dielectric function calculated by means of the effective-medium theory (EMT), first developed by Rytov in 1956 [35]. The theory is subject to the requirement that the thickness of the layers is much smaller than the wavelength of incident light. A simpler method for deriving the EMT, developed by Wood et al. [36], uses the characteristic matrix or T-matrix for transverse magnetic (TM) waves. This method considers incident plane waves and periodic boundary conditions, noting that wave vector components parallel to the layers remain unchanged. According to EMT, for a multilayer metamaterial with layers parallel to the x–y plane, the dielectric function components, in the limit of infinitesimal layer thickness, are given by:

$$\epsilon_{xx} = \epsilon_{yy} = \rho\epsilon_m + (1-\rho)\epsilon_d \tag{2}$$

$$\epsilon_{zz} = \frac{\epsilon_m \epsilon_d}{\rho\epsilon_d + (1-\rho)\epsilon_m} \tag{3}$$

where $\epsilon_m$ and $\epsilon_d$ are the dielectric functions of the metal and dielectric, respectively, and $\rho = t_m/(t_m + t_d)$ is the volume fraction of metal in the unit cell, which consists of a metal layer of thickness $t_m$ and a dielectric layer of thickness $t_d$. This is the first-order approximation, obtained by Taylor expanding the trigonometric functions in powers of layer thickness in the dispersion relation:

$$\begin{aligned}cos[k_z(t_m + t_d)] = & cos(k_m t_m)cos(k_d t_d) \\ & - \frac{1}{2}\left(\frac{\epsilon_d k_m}{\epsilon_m k_d} + \frac{\epsilon_d k_m}{\epsilon_m k_d}\right) sin(k_m t_m)sin(k_d t_d)\end{aligned} \tag{4}$$

where $k_{m,d}^2 = \epsilon_{m,d}\frac{\omega^2}{c^2} - k_z^2$. The requirement that the layer thickness is small compared to the wavelength of light means that the arguments of the cosine and sine functions in the above expression are small compared to unity; this justifies retaining only terms of low order in the expansion of these functions in powers of their arguments. Attempts were made to improve upon this approximation by taking higher-order terms in the expansion of Eq. (4) [37], using an operator formalism [38], or incorporating quantum optical effects [39].

Multilayer HMMs typically have unit cells on the scale of tens of nanometers [40]. The unit cells studied in this paper, however, are on the scale of tens of angstrom. Therefore, we also seek to test the accuracy of the EMT for layers on this scale.

## 2. Methods

The electronic structure calculations of the metal (BMO), the dielectric (SHO), and the metamaterial (BMO/SHO) are carried out using the augmented plane wave plus local orbitals method implemented in the Wien2k DFT code [41]. In this method, the space within the unit cell is divided into non-overlapping "muffin-tin" spheres centered at the nuclear sites, and the interstitial space between the spheres. The basis set for the expansion of the electronic wave function consists of atomic-like functions within the spheres, expanded in spherical harmonics up to $l_{max} = 10$, and plane waves in the interstitial region. The expansion in plane waves is such that $K_{max} R_{MT}^{min} = 7$, where $R_{MT}^{min}$ is the radius of the smallest muffin-tin sphere (in SHO, BMO, and BMO/SHO systems, this is the sphere centered on oxygen atom sites) and $K_{max}$ is the largest wave vector. The electron density in the interstitial region is also expanded in plane waves up to a maximum wave vector of $14/a_0$, with $a_0$ being the Bohr radius. Inside the spheres, the electron density is expanded in a product of radial functions and lattice harmonics with $l_{max} = 6$. Integration over the Brillouin zone is carried out using a mesh of $8 \times 8 \times 8$ k-points. The energy was converged to $10^{-4}$ Ry, while the charge was converged to $10^{-3}$ $e$.

The optical properties of the systems we consider in this work are calculated using the independent particle approximation (IPA), which disregards the electron–hole interaction experienced during absorption [42]. This approximation is well suited to metals, where the charge carriers are nearly free. We calculated that all materials examined in this paper except bulk SHO are metallic. Therefore, IPA could cause some error in the calculated dielectric function of bulk SHO, which we use in our analysis of the effective medium theory. However, we do not expect excitonic effects to change the dielectric functions appreciably.

Two contributions to the dielectric function $\epsilon(\omega)$ are considered: an intraband contribution, resulting from free electrons, and the interband one. For the intraband contribution, a Drude-like term is adopted:

$$\epsilon^{intra}(\omega) = 1 - \frac{\omega_p^2}{\omega^2 + i\omega\gamma} \quad (5)$$

where $\gamma$ is a damping constant and $\omega_p$ is the plasma frequency, which is $\sqrt{4\pi n e^2 / m^*}$ in cgs units, with $m^*$ being the effective electron mass and $n$ the free electron density [43]. For anisotropic materials, $\omega_p$ may be different for different directions in the crystal. This is because the effective mass is, similarly to the dielectric function, a second rank tensor, and its value depends on the direction of the electron's motion. This contribution is only significant for metals, because it requires free electrons.

In both metals and insulators, transitions of electrons from a state in one band to a state in another also contribute to the permittivity. An electron absorbs energy from a photon during the transitions, but because the wave vectors of photons at typical frequencies are very small compared to the Brillouin zone, these transitions are essentially vertical in a band diagram. The contribution of interband transitions to the imaginary part of the dielectric function is [42]:

$$Im[\epsilon_{ij}^{inter}(\omega)] = \frac{\hbar^2 e^2}{\pi m^2 \omega^2} \sum_{n,n'} \int_{\vec{k}} \langle n'\vec{k}|p_i|n\vec{k}\rangle \langle n\vec{k}|p_j|n'\vec{k}\rangle$$
$$\times \left(f(E_{n\vec{k}}) - f(E_{n'\vec{k}})\right) \delta(E_{n'\vec{k}} - E_{n\vec{k}} - \hbar\omega) \quad (6)$$

where $n$ and $n'$ are band indices, $f(E)$ is the Fermi distribution function, $p_i$ is the momentum operator in the $i$th direction, and $\delta(E)$ is the Dirac-delta function. The momentum matrix element, $\langle n\vec{k}|p_i|n'\vec{k}\rangle$, is proportional to the probability amplitude for a transition from a state with $n$ and $\vec{k}$ to one with $n'$ and $\vec{k}$ (again, because the momentum of the photon is negligible, $\vec{k}$ is approximately unchanged). The delta function is included due to conservation of energy; the final energy must be equal to the initial energy plus that of the photon. The Fermi occupation functions arise because transitions can only occur from an occupied state to an unoccupied state. Accurate calculation of the above expression requires a dense mesh of k-points. We therefore increased the mesh to $20 \times 20 \times 20$ for optical properties calculations. The real part of $\epsilon_{ij}(\omega)$ may be found from the imaginary part through the Kramers–Kronig relations. These may then be used to calculate optical conductivity or other optical properties.

To accurately calculate the optical properties of a dielectric, it is important that the calculation method yields an accurate value of the bandgap energy; otherwise, the calculated matrix elements which describe electronic transitions between valence and conduction bands would suffer large errors. It is well known that DFT calculations which approximate the exchange–correlation by semilocal functionals, such as in the local density approximation (LDA) or generalized gradient approximation (GGA), severely underestimate the bandgap energy. Improved values of the energy gap are obtained when hybrid functionals are used, though the resulting calculations are much more expensive in terms of time and computational resources. In these functionals, the exchange energy is a combination of a semilocal functional and a Hartree–Fock (HF) term. Typically, the HF fraction is 0.25, while the semilocal functional fraction is 0.75. Hence, in our calculations of the electronic and optical properties, we use hybrid functionals so as to describe correctly the electronic structure of the dielectric component of the metamaterials. For consistency, hybrid functionals with an HF fraction of 0.35 are used for all materials.

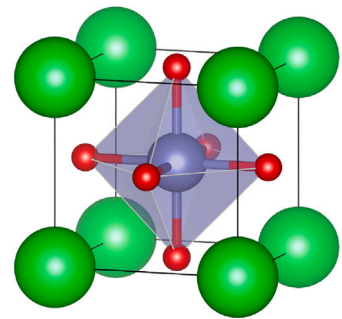

**Fig. 2.** The cubic unit cell of the ideal perovskite oxide $ABO_3$, with the A ions in green, B ion in purple, and oxygen ions in red.

## 3. Results and discussion

SrHfO$_3$ (SHO) and BaMoO$_3$ (BMO) are perovskite oxides, with a chemical formula of the form $ABO_3$. The unit cell is cubic, with A atoms at the cube corners, B atoms at the body center, and O atoms at the face centers; thus, each B atom sits at the center of a regular octahedron whose vertices are occupied by oxygen atoms (see Fig. 2). The lattice constant of SHO is 4.069 Å [44] and that of BMO is 4.0404 Å [45]. SHO is an insulator with a band gap of 6.1 eV [46], while BMO is metallic with a resistivity of 10–100 μΩ-cm [45].

DFT calculations on bulk SHO, using GGA and the Perdew–Burke–Ernzerhof (PBE) exchange–correlation energy [47] yield a bandgap of 3.87 eV. To produce the experimental value of the energy gap, we use a hybrid functional with a Hartree–Fock fraction given by 0.35; this raises the value of the bandgap to 6.01 eV. We considered spin–orbit coupling since Hf is relatively heavy, but its effect was not appreciable; the value of the band gap dropped by only 0.09 eV to 5.92 eV. Hence, in calculating the optical properties of the BMO/SHO metamaterials, the effect of spin–orbit coupling is not taken into account. In Fig. 3a, we plot the calculated electronic density of states of bulk SHO; the figure shows a wide bandgap of 6 eV in this dielectric material. On the other hand, DFT calculations show that bulk BMO is a metal, whether a PBE or hybrid functional is used. In Fig. 3b, we present the electronic density of states of bulk BMO, obtained using the hybrid functional. The finite density of states at the Fermi energy is indicative of the metallic nature of this material.

As indicated earlier, the use of gold or silver as the metal component in metamaterials entails considerable losses. In Fig. 4, we compare the observed and calculated $Im[\epsilon]$ of gold and silver with the calculated $Im[\epsilon]$ of BMO. We find that the experimental data give a range of values, with some equal to or lower than those of BMO in certain ranges. However, we believe it is most appropriate to draw conclusions by comparing DFT calculations with DFT calculations, using the same code and input parameters, as any errors are much more likely to be correlated for similar materials, as in the present case. Using this approach, the data shows that the imaginary part of the dielectric function of BMO is smaller than that of both gold and silver throughout the range of interest, providing evidence that losses in metamaterials using BMO as the plasmonic material could be reduced. Tuning of the optical properties of either or both constituent materials could result in further decreased losses, though more study would be necessary to determine the extent to which losses could be reduced.

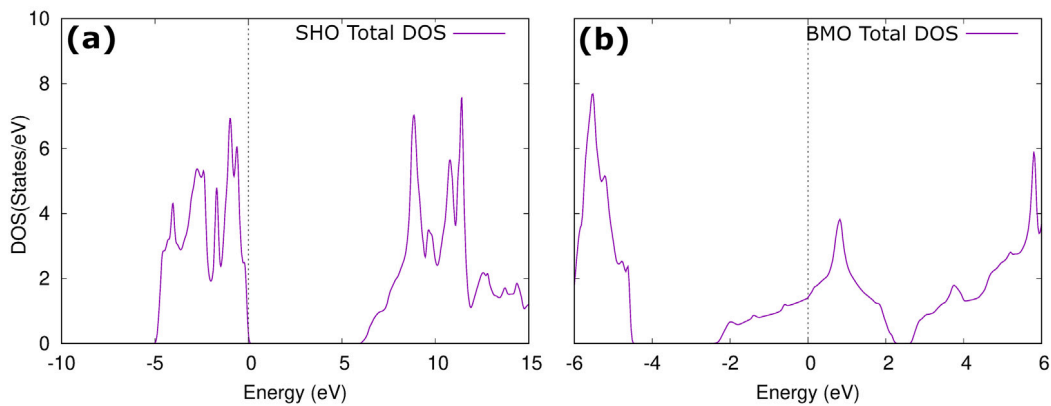

**Fig. 3.** Calculated densities of states for (a) SrHfO$_3$ and (b) BaMoO$_3$. For the case of SrHfO$_3$, the zero of energy is chosen to coincide with the top of the valence band, whereas for BaMoO$_3$, the Fermi energy is taken as zero.

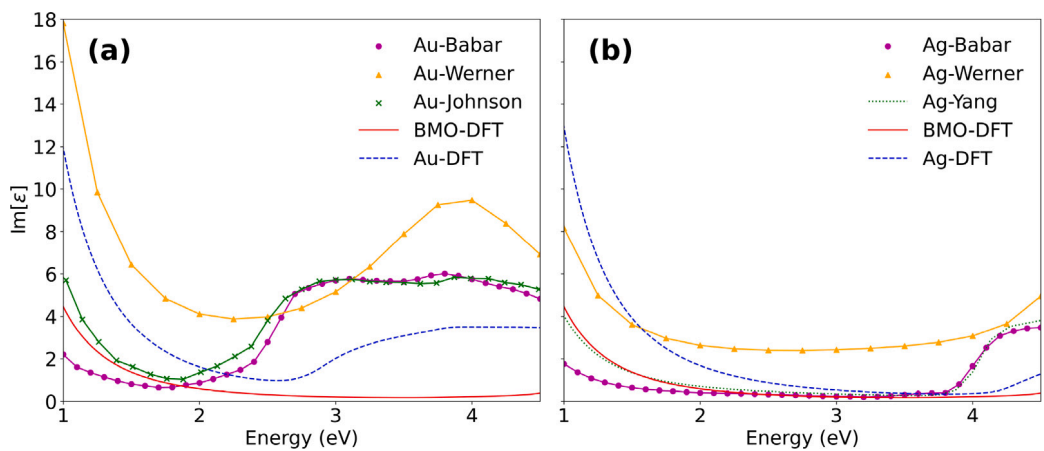

**Fig. 4.** (a) Imaginary parts of the dielectric functions of BMO obtained from DFT (solid red line) compared with gold using DFT calculations (broken blue line) and experimental values from Babar and Weaver [48] (magenta circles); Werner [49] (orange triangles); and Johnson and Christy [50] (green x). (b) BMO calculations compared with silver using DFT (broken blue line) and experimental values from Babar and Weaver (magenta circles), Werner (orange triangles), and Yang [51] (dotted green line).

The BMO/SHO multilayer metamaterial consists of a periodic stack of alternating metal BMO and dielectric SHO. In the calculations reported here, the unit cell of the SHO/BMO metamaterial consists of two unit cells of SHO (i.e., four layers: SrO, HfO$_2$, SrO, HfO$_2$), on top of which is one or more unit cells of BMO, corresponding to values of metal fraction $\rho$ ranging from 1/3 to 2/3. For example, for the case where the unit cell of the BMO/SHO metamaterial consists of one unit cell of BMO on top of the two unit cells of SHO, we have $\rho = 1/3$, while the case $\rho = 2/3$ corresponds to a unit cell of the metamaterial comprised of two SHO unit cells topped by four BMO unit cells. Once the unit cell is constructed, the atomic positions are relaxed. However, due to the similarity of the structures of SHO and BMO, relaxation of atomic positions does not lead to any significant deviations of atomic positions from the ideal ones.

The intraband portion of the dielectric function depends on the plasma frequency of the material. As mentioned in the methods section, anisotropy can lead to plasma frequency becoming a tensor. Because the construction of a layered metamaterial breaks symmetry along one direction, the plasma frequency of the materials under study have two unique components, $\omega_{p,xx}$ and $\omega_{p,zz}$. Respectively, the plasma frequencies $\omega_{p,xx}$ and $\omega_{p,zz}$ are (in eV) 3.288 and 0.084 for $\rho = 1/3$, 4.124 and 0.360 for $\rho = 1/2$, 4.190 and 0.067 for $\rho = 3/5$, and 4.392 and 0.094 and $\rho = 2/3$. The large difference between the plasma frequency components confirms that the material acts metallic in the plane and insulating out of the plane.

More evidence of this anisotropic behavior can be found in the band structure of these metamaterials. The energy bands along some high-symmetry directions in the Brillouin zone, for the case $\rho = 1/2$, are shown in Fig. 5b. It is clear that the bands along the $\Gamma$ to Z direction, corresponding to the c direction, are quite flat, indicating a large effective electron mass. This, in turn, indicates a poor electron mobility in that direction. Additionally, There are no bands crossing the Fermi level in this direction. One can then imagine that this material is semiconducting along that direction, with an out-of-plane band gap of about 1.8 eV. The corresponding density of states is shown in Fig. 5c. At the Fermi level, the DOS is dominated by states derived from the Mo d-orbitals, showing that states in the partially filled bands are derived mostly from these orbitals.

As can be seen in Fig. 6a, the calculated real part of the dielectric function of the $\rho = 1/3$ BMO/SHO metamaterial along the z-axis is negative, while the real part of the dielectric function in the xy-plane is positive up to about 2.4 eV, yielding a hyperbolic dispersion. The hyperbolic ranges for the other values of $\rho$ considered in this work extended up to more than 3 eV, as shown in Fig. 7. The infrared and visible ranges are those most important for technological applications, particularly for the hyperlens and nanolithography.

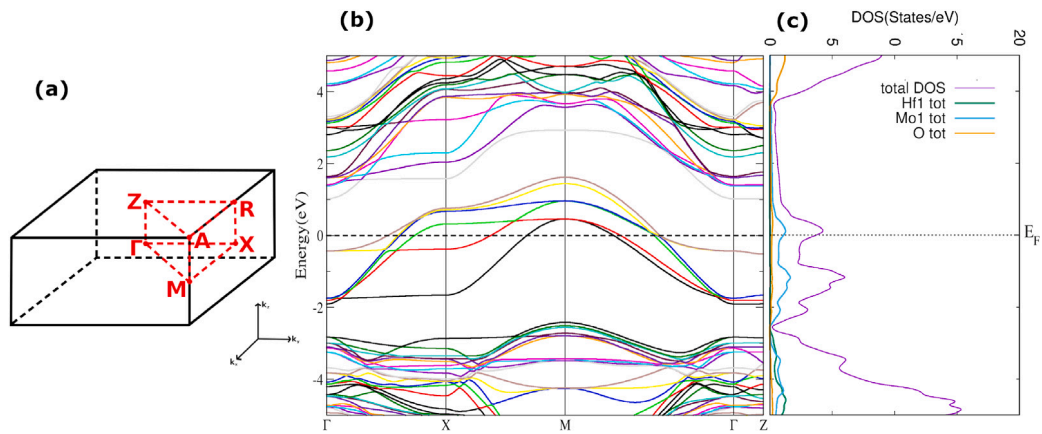

**Fig. 5.** The electronic properties of the $\rho = 1/2$ metamaterial. (a) The tetragonal 3d Brillouin zone, with high symmetry points labeled. (b) The band structure and (c) density of states, with the Fermi level set to 0 and marked with a dashed line. The colors of the bands have no significance. They are only intended to add clarity.

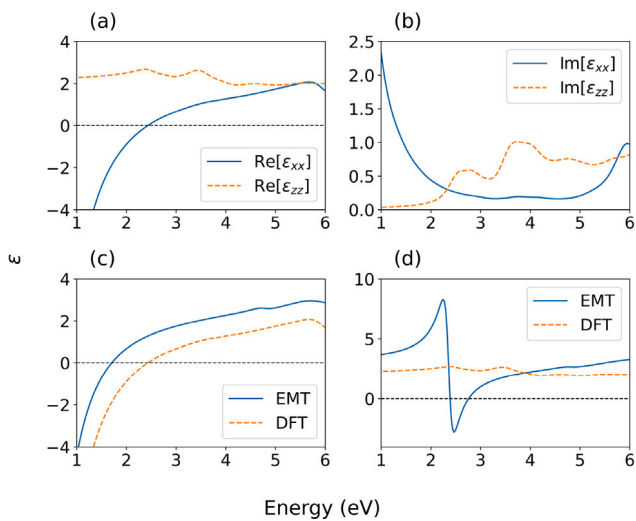

**Fig. 6.** The dielectric functions of the $\rho = 1/3$ HMM: (a) real parts from DFT (b) imaginary parts from DFT (c) real parts of $\epsilon_{xx}$ from DFT and EMT (d) real parts of $\epsilon_{zz}$ from EMT and DFT. $\epsilon = 0$ is marked with a dashed back line.

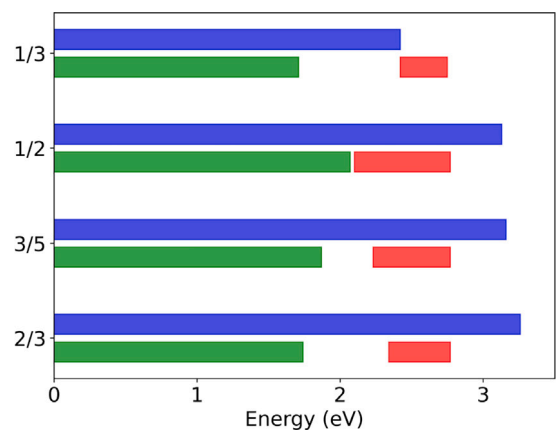

**Fig. 7.** The calculated hyperbolic ranges. The type 1 ranges predicted by DFT are shown in blue. The ranges predicted by the EMT of type 1 are shown in green, while type 2 are shown in red.

The optimal energy (or frequency) of a hyperbolic material is determined by the maximum of $-Re[\epsilon_{xx}(\omega)]/Im[\epsilon_{xx}(\omega)]$, which is sometimes referred to in the literature as the figure of merit (FoM) for hyperbolic materials [52]. This function increases when losses are smaller (lower $|Im[\epsilon]|$). The minus sign accounts for the fact that the $Re[\epsilon_{xx}]$ is negative in the type 1 hyperbolic range. For $\rho = 1/3, 1/2, 3/5$, and $2/3$, the maximum of the FoM occurs at 1.076, 1.784, 1.780, and 1.839 eV, respectively, falling in the visible range for all except $\rho = 1/3$. The corresponding wavelengths for these energies are 1152, 695, 696, and 674 nm, respectively.

The dielectric functions calculated from the EMT agreed with those calculated using DFT to some extent, but there were notable exceptions. The EMT predicted hyperbolic behavior in each case. However, it underestimated the type 1 hyperbolic range of the material, but remained within 1.5 eV of the DFT calculated range. In general, the EMT shifted the real part of dielectric functions upward, as can be seen in Fig. 6c for the $\rho = 1/3$ case. The results for the other values of $\rho$ were similar. For most of the energy range examined, the EMT produced dielectric function values larger than those calculated by DFT. Notable exceptions to this occurred at low energy (less than about 1.5 eV) and near points where the denominator in Eq. (3) approaches zero. You can see this latter behavior in Fig. 6d centered around 2.3 eV. This can produce a negative real part of $\epsilon_{zz}$, giving quite different behavior from DFT calculations. Significantly, the EMT indicates that each metamaterial should be type 2 hyperbolic in the visible range, while DFT predicts only type 1 hyperbolic behavior, as can be seen in Fig. 7. In the $\rho = 2/3$ and $3/5$ cases, both the real parts of $\epsilon_{xx}$ and $\epsilon_{zz}$ become negative for short block of frequencies, further shortening the type 1 hyperbolic range compared to DFT. Some difference between DFT and the EMT is expected, as the EMT is a purely classical treatment, while DFT is employed to solve quantum systems. In this case in particular, the layers are on the order of 1 nm, so we expect the quantum effects of confinement and tunneling between layers to contribute to the electronic properties.

It would be natural to expect that the imaginary part of the dielectric function would generally increase throughout the spectrum when the proportion of metal in the metamaterial is increased, since metals are associated with a much larger imaginary part, and therefore much larger losses. This generally matches what was found when calculating the dielectric functions with the EMT, with smaller fractions producing smaller values of $\epsilon$ at most points below about 7 eV. The difference between the largest and smallest value, however, was typically still less than one unit, and often much less. In our DFT calculations, on the other hand, the correlation between $\rho$ and the imaginary part of $\epsilon$ was only strong at smaller energies. Below about 2 or 3 eV, the relative size of each dielectric function was clearly tied to $\rho$, as can be seen in Fig. 8. Above this energy however, the correlation is very small or disappears altogether, and the imaginary part of the dielectric function for $\rho = 1/3$ is not particularly likely to be lower than any other fraction. This holds true for both $\epsilon_{xx}$ and $\epsilon_{zz}$.

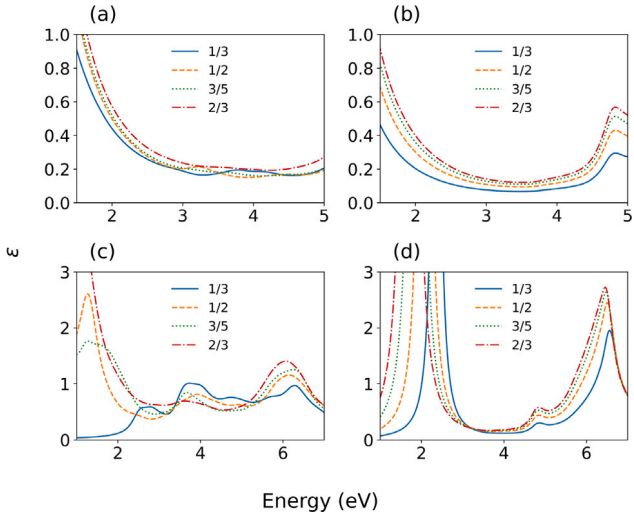

**Fig. 8.** Imaginary parts of the dielectric functions: (a) $\epsilon_{xx}$ from DFT, (b) $\epsilon_{xx}$ from the EMT, (c) $\epsilon_{zz}$ from DFT, and (d) $\epsilon_{zz}$ from the EMT. Each plot includes a line for all calculated values of $\rho$ (1/3, 1/2, 3/5, and 2/3).

We considered two potential corrections to the standard EMT. The non-local effects method includes terms that are second-order in layer thickness in the Taylor expansion of Eq. (4), yielding:

$$\epsilon_{zz}^{nl} = \frac{\epsilon_{zz}^0}{1 - \delta_{zz}}, \qquad \epsilon_{xx}^{nl} = \frac{\epsilon_{xx}^0}{1 - \delta_{xx}}$$

$$\delta_{zz} = \frac{t_m^2 t_d^2 (\epsilon_m - \epsilon_d) \epsilon_{zz}^{0\,2}}{12(t_m + t_d) \epsilon_d^2 \epsilon_m^2} \left( \epsilon_{xx}^0 \frac{\omega^2}{c^2} - \frac{k_x^2 (\epsilon_d + \epsilon_m)^2}{\epsilon_{xx}^{0\,2}} \right)$$

$$\delta_{xx} = \frac{t_m^2 t_d^2 (\epsilon_m - \epsilon_d) \epsilon_{zz}^{0\,2}}{12(t_m + t_d) \epsilon_{xx}^0} \frac{\omega^2}{c^2}$$

where $\epsilon_{zz}^0$ and $\epsilon_{xx}^0$ are the dielectric functions found from the standard EMT [37]. Another effective-medium approximation developed by Popov et al. uses an operator formalism, which leads to correction terms [38]:

$$\epsilon_{xx}^{op} = \epsilon_{xx}^0 + \frac{\omega^2 (t_m + t_d)^2}{6c^2} \sigma f(k_x) \widetilde{\epsilon}_{xx}$$

$$\epsilon_{zz}^{op} = \epsilon_{zz}^0 - \frac{\omega^2 (t_m + t_d)^2}{6c^2} \sigma \epsilon_{zz}^{0\,2} \left( \frac{2\rho - 1}{\epsilon_r} - \frac{f(k_x)}{\widetilde{\epsilon}_{zz}} \right)$$

$$\sigma = \rho(1-\rho)(\epsilon_d - \epsilon_m), \quad f(k_x) = \frac{k_x^2}{\frac{\omega^2}{c^2} \epsilon_r} - 1, \quad \epsilon_r = \left( \frac{1}{\epsilon_m} + \frac{1}{\epsilon_d} \right)^{-1}$$

$$\widetilde{\epsilon}_{xx} = \rho \epsilon_m - (1-\rho) \epsilon_d, \quad \widetilde{\epsilon}_{zz} = \left( \frac{\rho}{\epsilon_m} - \frac{1-\rho}{\epsilon_d} \right)^{-1}$$

In all of the above expressions, we are using a coordinate system such that $k_y = 0$.

We calculated the corrections to the real and imaginary parts of $\epsilon_{zz}$ and $\epsilon_{xx}$ for each value of $\rho$, considering two extreme cases: $k_x = 0$ and $k_z = 0$. Vanishing $k_x$ corresponds to light incident perpendicular to the layers. This simplifies the above expressions greatly. In this case, we find that the largest correction is 0.002 for $\text{Im}[\epsilon_{zz}(\omega)]$ with non-local effects and $\rho = 2/3$. When $k_z = 0$, we have light incident parallel to the layers. We can then use Eq. (1) to eliminate $k_x$, subject to the approximation that the $\epsilon_{zz}$ appearing in Eq. (1) is equal to $\epsilon_{zz}^0$. These corrections reach a maximum of 0.034 for $\text{Im}[\epsilon_{zz}(\omega)]$ with non-local effects and $\rho = 1/2$.

The correction terms are at least quadratic in layer thickness, which likely explains their very small values in the present case of layers on the order of 10 Å thick. We tested these same corrections with a theoretical unit cell thickness of 300 Å for $\rho = 2/3$, which is typical for realistic metamaterials [53,54]. In this case, the corrections to the dielectric functions reached a much larger maximum of 0.273 in $\text{Im}[\epsilon_{zz}(\omega)]$ with non-local effects for $k_x = 0$ and 1.905 in $\text{Re}[\epsilon_{zz}(\omega)]$ with non-local effects for $k_z = 0$. However, corrections of this size were rare. For the large majority of the calculated energy range (0 to 10 eV), corrections were smaller than 0.1. When plotted on top of the dielectric functions from the standard EMT, even those dielectric functions with the largest corrections (those using a theoretical 300 Å unit cell) are nearly indistinguishable, with only small visible differences at the extrema.

In conclusion, we have identified a plasmonic material with desirable characteristics in BMO, as well as metamaterials with hyperbolic dispersion in the technologically important visible and infrared ranges. Our calculations indicate that losses could be reduced in BMO/SHO metamaterials relative to those containing gold or silver, and there is the potential to further enhance loss reduction through substitution or other tuning methods. Additionally, BMO and SHO have the advantages of epitaxial growth and increased thermal stability. We also showed that the EMT agrees with DFT calculations in predicting type 1 hyperbolic behavior at low frequencies, although the EMT was inconsistent with DFT in its predictions of the type 1 hyperbolic frequency range and type 2 hyperbolic behavior at higher visible energies.

**Funding**

This work has been partially supported by the National Science Foundation, United States with Awards HRD-1547723 and HRD-2112554. The funding source had no involvement in any decisions regarding the research presented.

**CRediT authorship contribution statement**

**Jonathan Gjerde:** Writing – original draft, Investigation, Writing – review & editing, Visualization, Software. **Radi A. Jishi:** Conceptualization, Writing – review & editing, Resources, Investigation, Supervision.

**Declaration of competing interest**

The authors declare that they have no known competing financial interests or personal relationships that could have appeared to influence the work reported in this paper.

**Data availability**

Data will be made available on request.